\documentclass[showpacs,twocolumn,preprintnumbers,amsmath]{revtex4}

\usepackage{graphicx}
\usepackage{dcolumn}
\usepackage{bm}
\usepackage[]{hyperref}
\usepackage{setspace}
\usepackage{color}
\usepackage{overpic}

\newcommand{\sro}{{$^{88}$}Sr}

\newcommand{\tripl}{$^1$S$_0$-$^3$P$_1$}
\newcommand{\singl}{$^1$S$_0$-$^1$P$_1$}

\newcommand{\ws}{Wannier-Stark\,\,}

\draft

\begin{document}

\title{Coherent delocalization of atomic wavepackets in driven lattice
potentials}

\author{V. V. Ivanov, A. Alberti, M. Schioppo, G. Ferrari, M. Artoni$^{\S}$, M.
L. Chiofalo$^{\dag}$, G. M. Tino} \email{Guglielmo.Tino@fi.infn.it}

\affiliation{Dipartimento di Fisica and LENS - Universit\`a di Firenze, CNR-INFM,\\ INFN
- Sezione di Firenze, via Sansone 1, 50019 Sesto
Fiorentino, Italy \\
$^{\S}$ Department of Chemistry and Physics of Materials - University of Brescia, and
LENS, Italy \\
$^{\dag}$ CNISM and INFN, Classe di Scienze, Scuola Normale Superiore, Pisa, Italy}

\begin{abstract}
Atomic wavepackets loaded into a phase-modulated vertical optical-lattice potential
exhibit a coherent delocalization dynamics arising from \textit{intraband} transitions
among Wannier-Stark levels. Wannier-Stark intraband transitions are here observed by
monitoring the \emph{in situ} wavepacket extent. By varying the modulation frequency, we
find resonances at integer multiples of the Bloch frequency. The resonances show a
Fourier-limited width for interrogation times up to 2 seconds. This can also be used to
determine the gravity acceleration with ppm resolution.
\end{abstract}

\pacs{42.50.Vk, 03.75.Lm, 03.75.-b, 04.80.-y} \maketitle

Controlling quantum transport through an external driving field is a basic issue in
quantum-mechanics~\cite{grifoni98}, yet with relevance to fundamental physics tests and
precision measurements ~\cite{makhlin01} as well as to applications, such as the design
of novel miniaturized electronic~\cite{capasso90} and spintronic~\cite{wolf01} devices.
Quantum transport control has however gained a renewed interest with the advent of
optical lattices for ultracold atoms. These are in fact increasingly employed to realize
laboratory models for solid state crystals. The accurate tunability of atomic parameters
such as the temperature, the strength of interaction and the dimensionality, bring
ultracold atoms samples within the extreme quantum regime sought for precise quantum
transport control~\cite{bloch05}, gravity
measurements~\cite{anderson98,roati04,ferrari06bis}, and metrology~\cite{varenna06}.

Atoms transport control in optical lattices depends in general on the form of
the external driving field~\cite{gluck02} and, in particular, on its strength
and frequency whose values may be chosen so as to span from transport
enhancement~\cite{lin90} to suppression~\cite{grossmann91}. Within this context
Bloch oscillations~\cite{dahan96}, Landau-Zener tunnelling~\cite{wlkinson96},
and resonant tunnelling enhancement in tilted optical lattices~\cite{sias07}
are certainly worth being mentioned. Likewise important manifestations comprise
transport in the well-known kicked-atom model where quantum transport could
actually be engineered both by semiclassical \cite{sadgrove05} and by purely
quantum \cite{ryu06,behinaein06} effects.

In this Letter we experimentally demonstrate for the first time Wannier-Stark intraband
transitions in lattice potentials, a phenomenon which has been studied theoretically
\cite{Liu02,Thommen02} but has never been observed before. Our lattice potential has the
form:

\begin{equation}\label{Potential}
U(z,t)=m\hspace{1pt}g\hspace{1pt}z+\frac{U_0}{2} \cos\Big[2\hspace{1pt}k_L (z-z_0\cos (2
\pi \nu_M\,t)) \Big]\rule[-4mm]{0mm}{0mm}
\end{equation}
where $m\hspace{1pt}g\hspace{1pt}z$ is the gravity potential, $U_0$ is the lattice depth,
$k_L$ is the optical lattice wavevector while $z_0$ and $\nu_M$ are respectively the
phase-modulation amplitude and frequency.

\begin{figure}[t] \vspace{-2mm} \begin{center}
\hspace{-0mm}
\includegraphics[width=0.45\textwidth,angle=0]{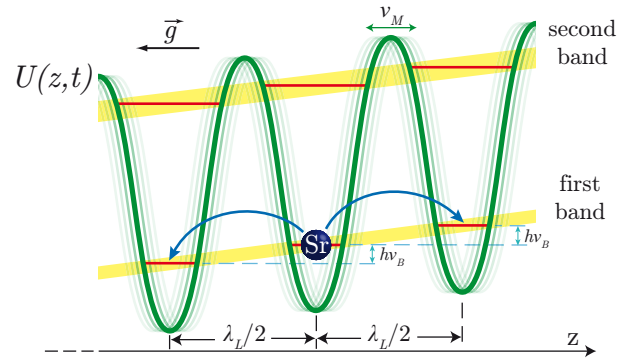}
\caption{\label{figura1}
Intraband site-to-site tunnelling resonantly sets in when the temporal modulation
frequency $\nu_M$ of the lattice phase is an integer multiple of the Bloch frequency
$\nu_B$ which corresponds to the potential energy between adjacent sites due to the
gravity acceleration $g$.}
\end{center}
\end{figure}

Intraband transitions between Wannier-Stark levels give rise to coherent
delocalization effects which we observe through a coherent ballistic expansion
of an initially well localized atomic wavepacket. Wannier-Stark intraband
tunneling, unlike the more familiar Landau-Zener tunnelling occurring between
different bands~\cite{wlkinson96,sias07}, is not affected by typical
decoherence mechanisms occurring in the Landau-Zener interband case, such as
line broadening due to the transverse profile of the lattice potential.
Furthermore we work with an atomic species remarkably robust against
decoherence processes \cite{StarkVlad,ferrari06}, which enables us to observe
transitions up to five neighboring Wannier-Stark levels, corresponding to a
coherently driven tunnelling across five neighboring sites. Owing to such a
quantum robustness the resonance spectra exhibit Fourier-limited widths over
excitation times of the order of seconds. Such a high-resolution enables us, in
turn, to measure the local acceleration of gravity with ppm relative precision.

We start by trapping and cooling about $2 \times 10^7$ \sro\, atoms at 3 mK in
a magneto-optical trap (MOT) operating on the \singl resonance line at 461 nm
\cite{ferrari06,ferrari06bis}. The temperature is further reduced by a second
cooling stage in a red MOT operating on the \tripl\, narrow transition at 689
nm. Finally we obtain $\sim 5 \times 10^5$ atoms at 1 $\mu$K. This preparation
phase takes about 2.5 s. Then, the red MOT is switched off and a
one-dimensional optical lattice is switched on adiabatically in 150 $\mu$s. The
lattice potential is originated by a single-mode frequency-doubled Nd:YVO$_4$
laser ($\lambda_L$ = 532 nm) delivering up to 170 mW on the atoms with a beam
waist of 100 $\mu$m. The beam is vertically aligned and retro-reflected by a
mirror producing a standing wave with a period $\lambda_L$/2 = 266 nm. The
corresponding photon recoil energy is $E_R = h^2 /2m \lambda_L^2 = k_B \times
381$ nK, and the maximum lattice depth is 20 $E_R$. In order  to modulate the
phase of the lattice potential, the retro-reflecting mirror is mounted on a
piezo-electric transducer (PZT) which is driven at frequency $\nu_M$ by a
synthesized frequency generator.

\newsavebox{\mysquare}\savebox{\mysquare}{\raisebox{0.4pt}
{\includegraphics[width=1.5mm]{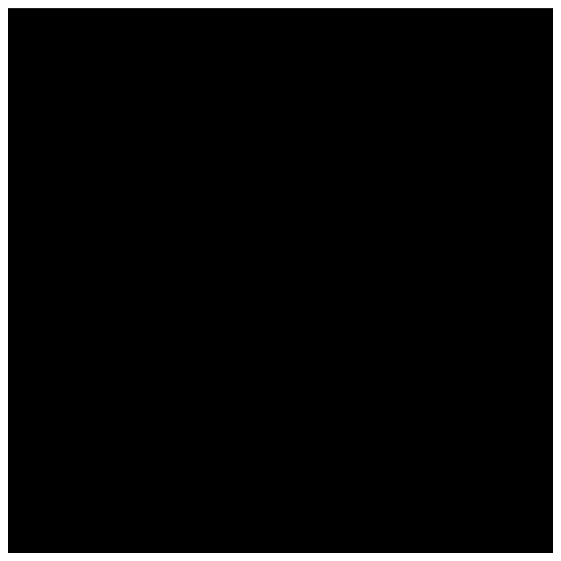}}}
\newsavebox{\mycircle}\savebox{\mycircle}{\raisebox{0.4pt}
{\includegraphics[width=1.5mm]{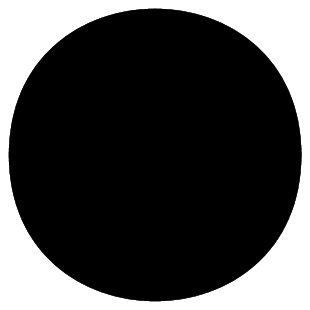}}}
\newsavebox{\mytriangle}\savebox{\mytriangle}{\raisebox{0.4pt}{\hspace
{-0.7pt}\includegraphics[width=2mm]{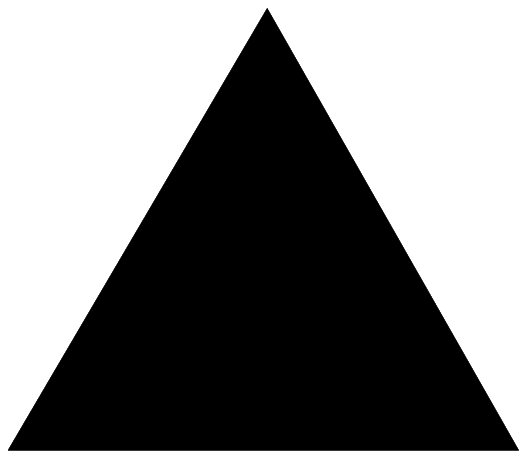}\hspace{-0.7pt}}}
\newsavebox{\mydiamond}\savebox{\mydiamond}{\raisebox{-0.4pt}{\hspace
{-0.7pt}\includegraphics[width=2mm]{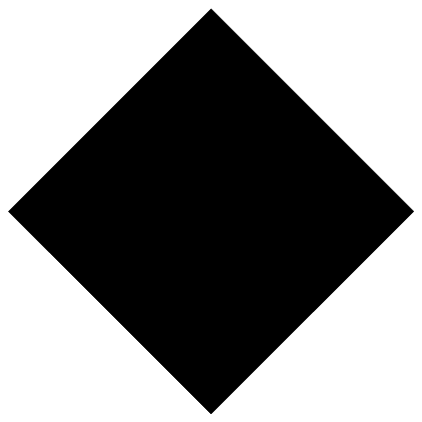}\hspace{-0.7pt}}}
\begin{figure}[t]
\scalebox{1}{\fboxsep=0cm\fboxrule=0.0pt\fbox{\begin{minipage}{1 \linewidth}
\vspace*{0mm} 
\begin{overpic}[width=0.85\textwidth,angle=0]{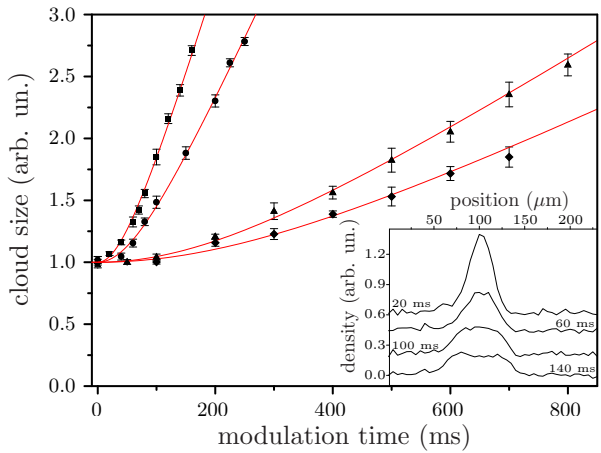}
\put(31,-5){modulation time (ms)} \put(-7,20){\rotatebox{90}{cloud size (arb. un.)}}
\put(53,8){\rotatebox{90}{\scalebox{0.75}{density (arb. un.)}}}
\put(73.,38.2){\scalebox{0.75}{position ($\mu$m)}}
\end{overpic}
\vspace*{5.5mm}
\end{minipage}}}
\begin{center}
\caption{\label{figura4} Wavepacket expansion as the lattice is modulated in phase at
four different frequencies $\nu_M=n \times \nu_B$, where (\usebox{\mysquare}) \emph{n}=1,
(\usebox{\mycircle}) \emph{n}=2, (\usebox {\mytriangle}) \emph{n}=3,
(\usebox{\mydiamond}) \emph{n}=4. The widths are normalized to the unperturbed initial
value $\sigma_0$. Inset: Atomic density profiles for the first (\emph{n=1}) harmonic
modulation at increasing modulation times (20 $\rightarrow$ 140 ms).}
\end{center}
\end{figure}

The voltage applied to the PZT allows to modulate the position of the lattice potential
by up to 6 sites peak-to-peak. The electronic-to-optical transfer function was verified
to be linear on the applied voltage and substantially independent from the considered
frequency. For a lattice potential depth corresponding to 20 $E_R$, the trap frequencies
are 71.5 kHz and 86 Hz in the longitudinal and and radial direction, respectively. Before
being transferred to the optical lattice, the atomic cloud in the red MOT has a disk
shape with a vertical size of 12 $\mu$m rms. In the transfer, the vertical extent is
preserved and we populate about 50 lattice sites with $10^5$ atoms. After letting the
atoms evolve in the optical lattice, we measure \emph{in situ} the spatial distribution
of the sample by absorption imaging of a resonant laser beam detected on a CCD camera.
The spatial resolution of the imaging system is 7 $\mu m$.

\begin{table}
 \caption{
 \label{BroadeningVelocity}
Root mean square broadening velocity and tunnelling rate of the confined atomic sample at
the different harmonics. The modulation depth is fixed to about 2 lattice sites
peak-to-peak, and the modulation frequency is resonant with the $n^{\rm th}$ harmonic
($\nu_M =  n \times \nu_B$). The thermal velocity in absence of the lattice potential is
10 mm/s rms.} \vspace{4mm}
 \begin{tabular}{lcccc}
   \hline\hline\\ [-2.0ex]
{\sc Resonance} ($\nu_M$/$\nu_B$) \hspace{8mm} &   1  &  2   &   3   &
 4    \\
\hline
                                &      &    & &        \\   [-1.5ex]
{\sc Expansion velocity} (mm/s) &  0.2\,\, & 0.13\,\, & 0.04\,\,  &  0.03\,\,
\\
{\sc Expansion velocity} (sites/s)  &  750 &  490 & 150 & 110
\\
{\sc Tunnelling rate} (s$^{-1}$)  &  750 &  245 & 50 & 27.5  \\
\hline\hline
        \\
 \end{tabular}
\end{table}

An atomic wavepacket moving in an optical lattice potential is characterized by an energy
and a quasi-momentum belonging to a specific band. Owing to the potential translational
symmetry, the wave-packet propagates typically unbound through the lattice. Under the
effect of a constant force $f_0$, however, the band splits into a series of \ws
resonances separated by integer multiples of the  Bloch frequency $\nu_{B}=\lambda_L
f_0/2 h$. In our case $f_0$ is the gravity force which breaks the translational symmetry
suppressing atomic tunneling between lattice sites, hence localizing the wavepacket, and
$\nu_B\approx575$ Hz. We observe indeed this localization in the absence of modulation
($z_0=0$) or for modulation frequencies $\nu_M$ far from \ws resonances. Conversely,
wavepacket delocalization, assessed through an increase of the atomic distribution width,
sets in instead for modulation frequencies $\nu_M=\nu_B$, as shown in the inset of Fig.
\ref{figura4}, or multiple integers of $\nu_B$ suggesting that tunnelling occurs not only
between nearest neighboring sites ($n=1$) but also between sites that are $n$~ lattice
periods apart ($n=2,3,4$). The atomic cloud spreads along the lattice axis and its width
is plotted  in Fig. \ref{figura4} for increasingly larger modulation times. At resonance
and after a transient due to the initial extent, the width grows linearly in time
undergoing a ballistic expansion as due to coherent site-to-site tunnelling. Broadening
proceeds at different velocities which we report in Tab. \ref{BroadeningVelocity}. These
are determined, for each $n$, by fitting the width with the function
$\sigma_n(t)=\sqrt{\sigma_0^2+v_n^2\, t^2}$, which is the convolution of two gaussians:
one accounting for the initial atomic distribution, the second accounting for the
wavepacket expansion. The delocalization slows-down with increasing $n$ due to a sharp
reduction of the tunnelling rate with increasing separation between the initial and final
\ws states. If $ v_n \simeq (n \lambda_L /2) \gamma_n$ is the wavepacket broadening
velocity associated with the $n-$th harmonic modulation, the relevant tunnelling rate
$\gamma_n$ across $n$ sites, as reported in Tab. \ref{BroadeningVelocity}, is seen to
decrease exponentially roughly as $3^{-n}$.

\begin{figure}[t]
\scalebox{1}{\fboxsep=0cm\fboxrule=0.0pt\fbox{\begin{minipage}{1 \linewidth}
\vspace*{0mm} 
\begin{overpic}[width=0.85\textwidth,angle=0]{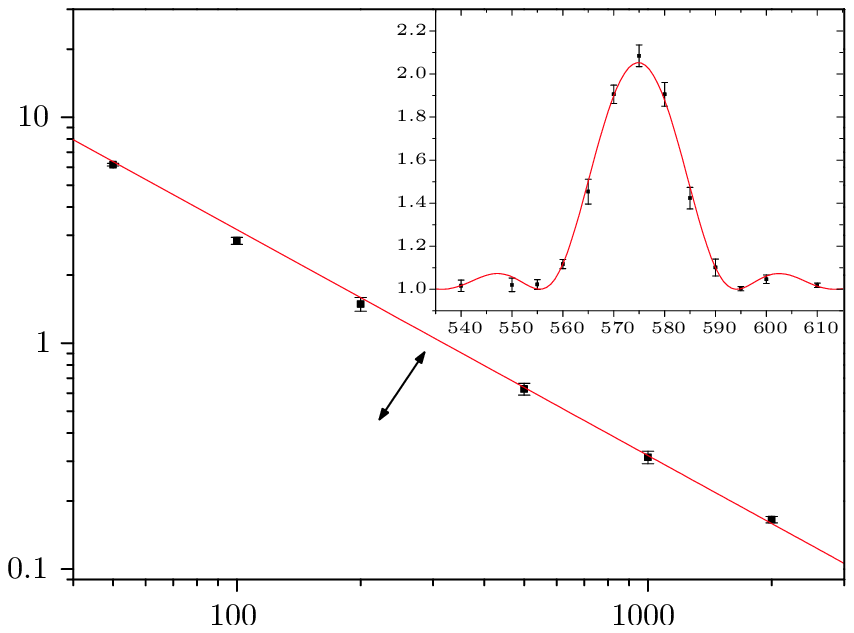}
\put(31,-5){modulation time (ms)} \put(-6,20){\rotatebox{90}{resonance width (Hz)}}
\put(42.5,42){\rotatebox{90}{\scalebox{0.7}{cloud size (arb. un.)}}}
\put(59,31){\scalebox{0.7}{modulation frequency (Hz)}}
\put(20,20){\scalebox{0.9}{$\displaystyle \Gamma(T)=\frac{1}{\pi\,T
(\textrm{s})}\;(\textrm{Hz})$}}
\end{overpic}
\vspace*{5.5mm}
\end{minipage}}}
\begin{center}
\caption{\label{figura2} Resonance width as function of the modulation time $T$. The
resonance is probed in the region $\nu_M \simeq \nu_B$. The line superposed to the data
point is the iperbola $(\pi\,T)^{-1}$ expected from a Fourier-limited resonance width in
a two level system. Inset: resonance spectrum for 50 ms excitation time. The fitting
function is of the form of Eq. \ref{SincFunction}.}
\end{center}
\end{figure}

While the dynamics of transitions between two distinct Wannier-Stark levels can be
described in terms of a generalized two level system, the spatial broadening, on the
other hand, can be ascribed to the iteration of the coherent tunnelling process over a
large number of lattice periods, typically 50 in the experiment. We verify this
hypothesis by studying the response of the system at different driving frequencies. First
we focus on the modulation close to the Bloch frequency $\nu_B$ and we study the shape of
the resonance. The inset of Fig. \ref{figura2} represents a typical data set of the
atomic extent for varying modulation frequencies, while keeping constant the amplitude of
modulation, the potential depth, and the excitation time. The data point are well fitted
with the function:
\begin{equation}\label{SincFunction}
    \sigma (\nu_M,t)=\sqrt{\sigma_0^2+ v^2_n\, t^2\, \, {\rm sinc}\left(
\frac{\nu_M-n\,\nu_B}{\Gamma}\right)^2}
\end{equation}
where $\sigma_0$ corresponds to the initial spatial extent, $v$ is the velocity of
spatial broadening at resonance, $t$ is the modulation time, ${\rm sinc}(x)$ is the
resonance function ${\rm sin}(x)/x$ for a two level transition probability and accounts
for the resonance term on the tunnelling rate, $n\,\nu_B$ is the resonance frequency, and
$\Gamma$ is the resonance half width at half maximum. The fit is remarkably good
supporting a model based on the generalized two level system.

\begin{figure}[t!]
\scalebox{1}{\fboxsep=0cm\fboxrule=0.0pt\fbox{\begin{minipage}{1 \linewidth}
\vspace*{0mm} 
\begin{overpic}[width=0.85\textwidth,angle=0]{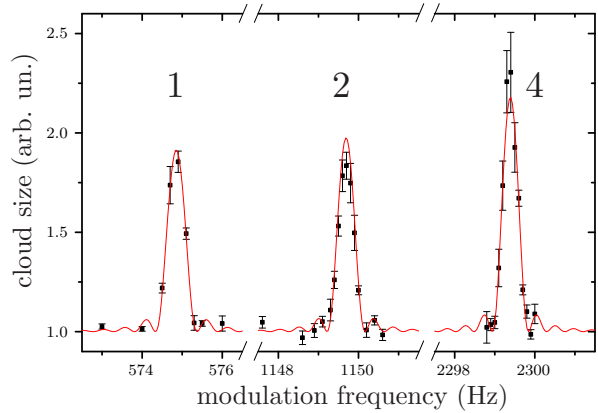}
\put(27.5,-5){modulation frequency (Hz)} \put(-6,16){\rotatebox{90}{cloud size (arb.
un.)}} \put(22,50.5){\scalebox{1.4}{1}} \put(52,50.5){\scalebox{1.4}{2}}
\put(87,50.5){\scalebox{1.4}{4}}
\end{overpic}
\vspace*{5.5mm}
\end{minipage}}}
\begin{center}
\caption{\label{figura3} Resonance spectra at the 1$^{\rm st}$, 2$^{\rm nd}$, and 4$^{\rm
th}$ harmonic of the Bloch frequency $\nu_B$. The excitation time is set to 2 seconds.
Within the error bars the fitted centerline frequencies are in integer multiple ratio.}
\end{center}
\end{figure}

We then measure the linewidth $\Gamma$ for different excitation times $T$ when modulating
at a frequency close to $\nu_B$ \cite{StarkModulation}. The results are plotted in Fig.
\ref{figura2} where we report the linewidth of the resonance at $\nu_B$ for an excitation
time varying between 50 ms to 2 s. The agreement of the datapoint with the superposed
hyperbola $1/(\pi \,T)$, expected from an ideal two level system, indicates that the
resonance linewidth is purely Fourier limited. Spurious incoherent processes may limit
the coherence time of the system on a timescale longer than 15 s, suggesting that the
delocalization dynamics is largely determined by coherent tunnelling. In fact, given the
initial size of the sample (12 $\mu m$ vertical extent equivalent to 50 lattice sites)
and the resolution of the imaging system, the driving induces a broadening of the atomic
distribution over a large number of lattice sites. If this were to be due to incoherent
tunnelling of the atoms between the lattice sites, such as in a random walk process, we
would expect a minimum resonance width equal to the Fourier limit multiplied by the total
number of jumps. A broadening over more than 50 lattice sites, as we observe, would yeald
a resonance linewidth orders of magnitude larger than the one we observe in the
experiment. In case of a random walk in the lattice sites, at long times we would further
expect a spatial broadening increasing as the square root of the time, again this is not
consistent with our observations.

These Fourier-limited resonances turn out to be a powerful tool to measure accelerations
with high accuracy. Similarly to the first harmonic, higher harmonics also exhibit a
Fourier limited resonance linewidth for an interaction time longer than 2 s. In Fig.
\ref{figura3} we compare the resonance shape at $\nu_B$ with those at $2 \nu_B$ and $4
\nu_B$ for a 2 s excitation time \cite{StarkModulation}. The different resonances are
quite similar in shape and within the error bars we find that they remain Fourier limited
regardless of the order of the harmonic. Previous applications of Bloch frequency
measurements to determine the gravity acceleration had a resolution limited by the
quality factor $\nu_B/\delta\nu$ of the line (where $\delta\nu$ was the Fourier limit set
either by the coherence time  \cite{anderson98}, or the lifetime of the sample
\cite{ferrari06bis}), and by the signal-to-noise ratio which depended also on how much
the temperature of the atoms is lower than the recoil energy. It is worth noting that in
our case the initial temperature is about twice the recoil.

Our results can be exploited to improve acceleration measurements resolution owing to the
absence of a specific requirement on the sample temperature with respect to the photon
recoil energy and to the possibility of measuring higher harmonics of $\nu_B$ at a
constant resonance linewidth (see Fig. \ref{figura3}). Working with atoms at a
temperature nearly at or above the recoil temperature reduces substantially the technical
constraints on sample preparation, making more atoms available in the test sample, and
making possible the employ of additional atomic or molecular species which can not be
cooled to sub-recoil temperatures. In addition, working at higher harmonics with a
constant resonance linewidth allows to improve the line quality factor by the index of
the considered harmonic. This improves the final resolution on the acceleration
measurement correspondingly. Modulating over 2 s, we measured $\nu_B=(574.8459 \pm
0.0015)$ Hz, which yields a local gravity acceleration g$=(9.805301\pm 0.000026)$
m/s$^{2}$\cite{Verticality}. This resolution of 2 ppm, which improves the previous
state-of-the-art by a factor 3 \cite{ferrari06bis}, is limited by the 1 s
background-limited lifetime of our vacuum system. Minor modifications of the experimental
apparatus should allow an improvement of the sensitivity by at least one order of
magnitude.

Delocalization of cold atoms wavepackets in a periodically driven optical lattice
occurring through coherent intraband tunneling is here thoroughly investigated. Control
over such delocalization enables us to modify the atoms wavefunction extent over regions
that are about 50 times their thermal de Broglie wavelength stretching, as in our case,
the initial 200 nm wavepacket width to more than $10\, \mu m$. Under our experimental
conditions the wavepacket expansion increases linearly with the lattice modulation
amplitude, though possible nonlinearities in the response may arise and will be the
object of future investigations. Coherent intraband resonant tunneling turns out to be
quite practical for increasing the sensitivity of force measurements with sub-millimeter
spatial resolution as in the case of Casimir forces and in Newtonian gravity at small
distances \cite{ferrari06bis}. It may also be useful for atomtronic devices such as
parallel quantum atomic couplers.

We thank G. C. La Rocca for fruitful discussions, F. S. Pavone for the lending
of part of the equipment, and R. Ballerini, M. De Pas, M. Giuntini, A. Hajeb,
A. Montori for technical assistance. This work was supported by LENS, INFN, EU
(under contract RII3-CT-2003 506350 and the FINAQS project), ASI and Ente CRF.


\begin{thebibliography}{29}
\expandafter\ifx\csname natexlab\endcsname\relax\def\natexlab#1{#1}\fi
\expandafter\ifx\csname bibnamefont\endcsname\relax
  \def\bibnamefont#1{#1}\fi
\expandafter\ifx\csname bibfnamefont\endcsname\relax
  \def\bibfnamefont#1{#1}\fi
\expandafter\ifx\csname citenamefont\endcsname\relax
  \def\citenamefont#1{#1}\fi
\expandafter\ifx\csname url\endcsname\relax
  \def\url#1{\texttt{#1}}\fi
\expandafter\ifx\csname urlprefix\endcsname\relax\def\urlprefix{URL }\fi
\providecommand{\bibinfo}[2]{#2} \providecommand{\eprint}[2][]{\url{#2}}


\bibitem[{\citenamefont{Grifoni and Hanggi}(1998)}]{grifoni98}
\bibinfo{author}{\bibfnamefont{M.}~\bibnamefont{Grifoni}} \bibnamefont{and}
  \bibinfo{author}{\bibfnamefont{P.}~\bibnamefont{Hanggi}},
  \bibinfo{journal}{Phys. Rep.} \textbf{\bibinfo{volume}{304}},
  \bibinfo{pages}{229} (\bibinfo{year}{1998}).


\bibitem[{\citenamefont{Makhlin et~al.}(2001)\citenamefont{Makhlin, Schon, and
  Shnirman}}]{makhlin01}
\bibinfo{author}{\bibfnamefont{Y.}~\bibnamefont{Makhlin}},
  \bibinfo{author}{\bibfnamefont{G.}~\bibnamefont{Schon}}, \bibnamefont{and}
  \bibinfo{author}{\bibfnamefont{A.}~\bibnamefont{Shnirman}},
  \bibinfo{journal}{Rev. Mod. Phys.} \textbf{\bibinfo{volume}{73}},
  \bibinfo{pages}{357} (\bibinfo{year}{2001}).


  \bibitem[{\citenamefont{Capasso and Datta}(1990)}]{capasso90}
\bibinfo{author}{\bibfnamefont{F.}~\bibnamefont{Capasso}} \bibnamefont{and}
  \bibinfo{author}{\bibfnamefont{S.}~\bibnamefont{Datta}},
  \bibinfo{journal}{Physics Today} \textbf{\bibinfo{volume}{43}},
  \bibinfo{pages}{74} (\bibinfo{year}{1990}).


\bibitem[{\citenamefont{Wolf et~al.}(2001)\citenamefont{Wolf, Awschalom, and
  {Buhrman {\it et al.}}}}]{wolf01}
\bibinfo{author}{\bibfnamefont{S.~A.} \bibnamefont{Wolf {\it et al.}}},
  \bibinfo{journal}{Science} \textbf{\bibinfo{volume}{294}},
  \bibinfo{pages}{1488} (\bibinfo{year}{2001}).
  \bibinfo{author}{\bibfnamefont{B.~T.} \bibnamefont{Seaman}},
  \bibinfo{author}{\bibfnamefont{M.}~\bibnamefont{Kramer}},
  \bibinfo{author}{\bibfnamefont{D.~Z.} \bibnamefont{Anderson}},
  \bibnamefont{and} \bibinfo{author}{\bibfnamefont{M.~J.}
  \bibnamefont{Holland}}, \bibinfo{journal}{Phys. Rev. A}
  \textbf{\bibinfo{volume}{75}}, \bibinfo{pages}{23615} (\bibinfo{year}{2007}).



\bibitem[{\citenamefont{Bloch}(2005, and references therein)}]{bloch05}
\bibinfo{author}{\bibfnamefont{I.}~\bibnamefont{Bloch}}, \bibinfo{journal}{Nat.
  Phys.} \textbf{\bibinfo{volume}{1}}, \bibinfo{pages}{253001}
  (\bibinfo{year}{2005}), and references therein.


\bibitem[{\citenamefont{Anderson and Kasevich}(1998)}]{anderson98}
\bibinfo{author}{\bibfnamefont{B.~P.} \bibnamefont{Anderson}} \bibnamefont{and}
  \bibinfo{author}{\bibfnamefont{M.~A.} \bibnamefont{Kasevich}},
  \bibinfo{journal}{Science} \textbf{\bibinfo{volume}{282}},
  \bibinfo{pages}{1686} (\bibinfo{year}{1998}).

\bibitem[{\citenamefont{Roati et~al.}(2004)\citenamefont{Roati, de~Mirandes,
  Ferlaino, Ott, Modugno, and Inguscio}}]{roati04}
\bibinfo{author}{\bibfnamefont{G.}~\bibnamefont{Roati {\it et al.}}},
  \bibinfo{journal}{Phys. Rev. Lett.} \textbf{\bibinfo{volume}{92}},
  \bibinfo{pages}{230402} (\bibinfo{year}{2004}).

\bibitem[{\citenamefont{Ferrari
  et~al.}(2006{\natexlab{a}})\citenamefont{Ferrari, Poli, Sorrentino, and
  Tino}}]{ferrari06bis}
\bibinfo{author}{\bibfnamefont{G.}~\bibnamefont{Ferrari}},
  \bibinfo{author}{\bibfnamefont{N.}~\bibnamefont{Poli}},
  \bibinfo{author}{\bibfnamefont{F.}~\bibnamefont{Sorrentino}},
  \bibnamefont{and} \bibinfo{author}{\bibfnamefont{G.}~\bibnamefont{Tino}},
  \bibinfo{journal}{Phys. Rev. Lett.} \textbf{\bibinfo{volume}{97}},
  \bibinfo{pages}{060402} (\bibinfo{year}{2006}{\natexlab{a}}).


\bibitem[{\citenamefont{Haensch et~al.}(2006)\citenamefont{Haensch, Leschiutta,
  and Wallard}}]{varenna06}
\bibinfo{editor}{\bibfnamefont{T.}~\bibnamefont{H{\"a}nsch}},
  \bibinfo{editor}{\bibfnamefont{S.}~\bibnamefont{Leschiutta}},
  \bibnamefont{and} \bibinfo{editor}{\bibfnamefont{A.}~\bibnamefont{Wallard}},
  eds., \emph{\bibinfo{title}{{\it Metrology and Fundamental Constants}, {\rm
  Proceedings of the International School of Physics "Enrico Fermi"}}}
  (\bibinfo{publisher}{IOS Press, Amsterdam}, \bibinfo{year}{2006}).

\bibitem[{\citenamefont{gluck02}(2002)}]{gluck02}
\bibinfo{author}{\bibfnamefont{M.}~\bibnamefont{Gl{\"u}ck}},
\bibinfo{author}{\bibfnamefont{A.~R.} \bibnamefont{Kolovsky}} \bibnamefont{and}
  \bibinfo{author}{\bibfnamefont{H.~J.} \bibnamefont{Korsch}},
  \bibinfo{journal}{Phys. Rep.} \textbf{\bibinfo{volume}{366}},
  \bibinfo{pages}{103} (\bibinfo{year}{2002}).


\bibitem[{\citenamefont{Lin and Ballentine}(1990)}]{lin90}
\bibinfo{author}{\bibfnamefont{W.~A.} \bibnamefont{Lin}} \bibnamefont{and}
  \bibinfo{author}{\bibfnamefont{L.~E.} \bibnamefont{Ballentine}},
  \bibinfo{journal}{Phys. Rev. Lett.} \textbf{\bibinfo{volume}{65}},
  \bibinfo{pages}{2927} (\bibinfo{year}{1990}).


\bibitem[{\citenamefont{Grossmann et~al.}(1991)\citenamefont{Grossmann,
  Dittrich, Jung, and Hanggi}}]{grossmann91}
\bibinfo{author}{\bibfnamefont{F.}~\bibnamefont{Grossmann}},
  \bibinfo{author}{\bibfnamefont{T.}~\bibnamefont{Dittrich}},
  \bibinfo{author}{\bibfnamefont{P.}~\bibnamefont{Jung}}, \bibnamefont{and}
  \bibinfo{author}{\bibfnamefont{P.}~\bibnamefont{H{\"a}nggi}},
  \bibinfo{journal}{Phys. Rev. Lett.} \textbf{\bibinfo{volume}{67}},
  \bibinfo{pages}{516} (\bibinfo{year}{1991}).


\bibitem[{\citenamefont{Dahan et~al.}(1996)\citenamefont{Dahan, Peik, Reichel,
  Castin, and Salomon}}]{dahan96}
\bibinfo{author}{\bibfnamefont{M.~B.} \bibnamefont{Dahan}},
  \bibinfo{author}{\bibfnamefont{E.}~\bibnamefont{Peik}},
  \bibinfo{author}{\bibfnamefont{J.}~\bibnamefont{Reichel}},
  \bibinfo{author}{\bibfnamefont{Y.}~\bibnamefont{Castin}}, \bibnamefont{and}
  \bibinfo{author}{\bibfnamefont{C.}~\bibnamefont{Salomon}},
  \bibinfo{journal}{Phys. Rev. Lett.} \textbf{\bibinfo{volume}{76}},
  \bibinfo{pages}{4508} (\bibinfo{year}{1996}).


\bibitem[{\citenamefont{Wilkinson et~al.}(1996)\citenamefont{Wilkinson,
  Bharucha, Madison, Niu, and Raizen}}]{wlkinson96}
\bibinfo{author}{\bibfnamefont{S.~R.} \bibnamefont{Wilkinson}},
  \bibinfo{author}{\bibfnamefont{C.~F.} \bibnamefont{Bharucha}},
  \bibinfo{author}{\bibfnamefont{K.~W.} \bibnamefont{Madison}},
  \bibinfo{author}{\bibfnamefont{Q.}~\bibnamefont{Niu}}, \bibnamefont{and}
  \bibinfo{author}{\bibfnamefont{M.~G.} \bibnamefont{Raizen}},
  \bibinfo{journal}{Phys. Rev. Lett.} \textbf{\bibinfo{volume}{76}},
  \bibinfo{pages}{4512} (\bibinfo{year}{1996}).



\bibitem[{\citenamefont{Sias et~al.}(2007)\citenamefont{Sias, Zenesini,
  Lignier, Wimberger, Ciampini, Morsch, and Arimondo}}]{sias07}
\bibinfo{author}{\bibfnamefont{C.}~\bibnamefont{Sias {\it et al.}}},
  \bibinfo{journal}{Phys. Rev. Lett.} \textbf{\bibinfo{volume}{98}},
  \bibinfo{pages}{120403} (\bibinfo{year}{2007}).

\bibitem[{\citenamefont{Sadgrove et~al.}(2005)\citenamefont{Sadgrove,
  Wimberger, Parkins, and Leonhardt}}]{sadgrove05}
\bibinfo{author}{\bibfnamefont{M.}~\bibnamefont{Sadgrove}},
  \bibinfo{author}{\bibfnamefont{S.}~\bibnamefont{Wimberger}},
  \bibinfo{author}{\bibfnamefont{S.}~\bibnamefont{Parkins}}, \bibnamefont{and}
  \bibinfo{author}{\bibfnamefont{R.}~\bibnamefont{Leonhardt}},
  \bibinfo{journal}{Phys. Rev. Lett.} \textbf{\bibinfo{volume}{94}},
  \bibinfo{pages}{174103} (\bibinfo{year}{2005}).

\bibitem[{\citenamefont{Ryu et~al.}(2006)\citenamefont{Ryu, Andersen, Vaziri,
  d'Arcy, Grossman, Helmerson, and Phillips}}]{ryu06}
\bibinfo{author}{\bibfnamefont{C.}~\bibnamefont{Ryu {\it et al.}}},
 \bibinfo{journal}{Phys. Rev. Lett.}
  \textbf{\bibinfo{volume}{96}}, \bibinfo{pages}{160403}
  (\bibinfo{year}{2006}).

\bibitem[{\citenamefont{Behinaein et~al.}(2006)\citenamefont{Behinaein,
  Ramareddy, Ahmadi, and Summy}}]{behinaein06}
\bibinfo{author}{\bibfnamefont{G.}~\bibnamefont{Behinaein}},
  \bibinfo{author}{\bibfnamefont{V.}~\bibnamefont{Ramareddy}},
  \bibinfo{author}{\bibfnamefont{P.}~\bibnamefont{Ahmadi}}, \bibnamefont{and}
  \bibinfo{author}{\bibfnamefont{G.~S.} \bibnamefont{Summy}},
  \bibinfo{journal}{Phys. Rev. Lett.} \textbf{\bibinfo{volume}{97}},
  \bibinfo{pages}{244101} (\bibinfo{year}{2006}).

\bibitem[{\citenamefont{Glueck et~al.}(2000)\citenamefont{Glueck, Hankel,
  Kolovsky, and Korsch}}]{Liu02}
  \bibinfo{author}{\bibfnamefont{W.~M.} \bibnamefont{Liu {\it et al.}}},
  \bibinfo{journal}{Phys. Rev. Lett.} \textbf{\bibinfo{volume}{88}},
  \bibinfo{pages}{170408} (\bibinfo{year}{2002}).

\bibitem[{\citenamefont{Glueck et~al.}(2000)\citenamefont{Glueck, Hankel,
  Kolovsky, and Korsch}}]{Thommen02}
  \bibinfo{author}{\bibfnamefont{Q.}~\bibnamefont{Thommen}},
  \bibinfo{author}{\bibfnamefont{J.~C.} \bibnamefont{Garreau}},
  \bibnamefont{and} \bibinfo{author}{\bibfnamefont{V.}~\bibnamefont{Zehnl\'e}},
  \bibinfo{journal}{Phys. Rev. A} \textbf{\bibinfo{volume}{65}},
  \bibinfo{pages}{053406} (\bibinfo{year}{2002}).

\bibitem[{Sta({\natexlab{a}})}]{StarkVlad}
\emph{\bibinfo{title}{{\rm Atomic $^{88}$Sr in the ground state is a zero-spin
  particle, hence not sensitive to magnetic fields, and has a negligible
  elastic cross section, which prevents the loss of coherence from atom-atom
  interactions.}}}

\bibitem[{\citenamefont{Ferrari
  et~al.}(2006{\natexlab{b}})\citenamefont{Ferrari, Drullinger, Poli,
  Sorrentino, and Tino}}]{ferrari06}
\bibinfo{author}{\bibfnamefont{G.}~\bibnamefont{Ferrari}},
  \bibinfo{author}{\bibfnamefont{R.~E.} \bibnamefont{Drullinger}},
  \bibinfo{author}{\bibfnamefont{N.}~\bibnamefont{Poli}},
  \bibinfo{author}{\bibfnamefont{F.}~\bibnamefont{Sorrentino}},
  \bibnamefont{and} \bibinfo{author}{\bibfnamefont{G.}~\bibnamefont{Tino}},
  \bibinfo{journal}{Phys. Rev. A} \textbf{\bibinfo{volume}{73}},
  \bibinfo{pages}{23408} (\bibinfo{year}{2006}{\natexlab{b}}).

\bibitem[{Sta({\natexlab{b}})}]{StarkModulation}
\emph{\bibinfo{title}{{\rm For each data set the amplitude of the phase
  modulation is chosen in order to double the spatial extent of the sample at
  resonance.}}}


\bibitem[{ver({\natexlab{b}})}]{Verticality}
\emph{\bibinfo{title}{{\rm The green trapping beam was initially aligned along the
vertical direction with an accuracy better than 1 mrad, resulting in an accuracy better
than 1 ppm. Following misalignments may affect the accuracy of the measurement but not
its resolution.}}}

\end{thebibliography}
\end{document}